\documentclass[twocolumn,prl]{revtex4}
\usepackage{graphics,graphicx}
\begin{document}
\title{Unusual Coupling Between Field-induced Spin Fluctuations and Spin Density Wave in Intermetallic CeAg$_{2}$Ge$_{2}$}
\author{D.K.~Singh$^{1,2}$}
\author{A.~Thamizhavel$^{3}$}
\author{S.~Chang$^{1}$}
\author{J.W.~Lynn$^{1}$}
\author{D.A.~Joshi$^{3}$}
\author{S.K.~Dhar$^{3}$}
\author{S.~Chi$^{1,2}$}
\affiliation{$^{1}$NIST Center for Neutron Research, Gaithersburg, MD 20899, USA}
\affiliation{$^{2}$Department of Materials Science and Engineering, University of Maryland, College Park, MD 20742,USA}
\affiliation{$^{3}$Tata Institute of Fundamental Research, Mumbai, INDIA}

\begin{abstract}
We report on experimental evidence for an unusual coupling between the magnetic field-induced fluctuations of correlated Ce-ions coinciding with the discontinuous movement of the underlying spin density wave in the intermetallic rare earth compound CeAg$_{2}$Ge$_{2}$. The measurements performed using neutron scattering and magnetic Gruneisen ratio methods suggest that the coupling onsets at $H$= 2.7 T, $T$$\leq$ 3.8 K and persists to the lowest measurement temperature $T$$\simeq$0.05 K. These measurements suggest a new mechanism behind the spin fluctuations which can affect the intrinsic properties of the system.
\end{abstract}

 \maketitle

Intermetallic rare-earth compounds containing a lattice of 4$f$ or 5$f$-electrons (such as Ce, U and Yb) are prototypical systems to study the magnetic quantum phase transition (QPT); a second order phase transition separating the paramagnetic state from the magnetically ordered state at $T$=0 K.\cite{Steglich,Si,Lohneysen,Coleman} The standard model of QPT based on the Hertz-Millis-Moriya spin-fluctuation theory invokes the idea that the QPT results from the fluctuations of the antiferromagnetic moment (order parameter) and their intensity diverges at the QPT.\cite{Hertz,Millis,Moriya} The dynamic behavior, manifested by spin fluctuations, is experimentally probed using an external control parameter such as pressure, chemical doping or magnetic field. While the synergistic efforts of theoretical and experimental investigations have led to a broader understanding of the mechanism behind spin fluctuations and the related scaling of the dynamic spin susceptibility, new behaviors have also emerged especially in itinerant magnetic systems which prohibit a universal formulation and require further experimental investigations of spin fluctuation phenomena.\cite{Knafo,Kadowaki,Aronson,Stockert,Wilson}

In this article, we report on the experimental investigation of a Ce-based intermetallic compound, CeAg$_{2}$Ge$_{2}$ which belongs to I4/mmm tetragonal space group with lattice constants of a = b = 4.305 $\AA$ and c = 10.971 $\AA$. The study of CeAg$_{2}$Ge$_{2}$ is interesting in itself as it exhibits an interplay of modest heavy Fermion behavior and long range antiferromagnetic order, $T_N = 4.6$~K.\cite{Loidl,Thamizhavel} The paramagnetic susceptibility shows an easy [100] direction in the tetragonal plane and an effective magnetic moment per Ce-ion of 2.54 $\mu$$_{B}$. The lack of single crystals until now has inhibited the determination of the ground state magnetic configuration and the detailed exploration of possible critical behavior in this compound. We have performed a series of neutron scattering and magnetic Gruneisen ratio measurements on single crystal CeAg$_{2}$Ge$_{2}$ in applied magnetic field up to 9 T. The ground state magnetic configuration of correlated Ce-ions forms a spin density wave (SDW) magnetic structure with Ce-spins fluctuating spatially along the $b$-axis (equivalent to $a$-axis). However, the main result of this paper is the discovery of an unusual coupling between the moment fluctuations of correlated Ce-ions, also manifested by a subtle change in the ordered moment, and the discontinuous movement of the underlying SDW configuration in applied magnetic fields along the [100] direction. It is found that while the spins fluctuate temporally along the [0K0] direction, collectively they move along the [00L] direction in a discontinuous way at $H$ = 2.7 T and $T$$\leq$3.8 K. The discontinuous movement of correlated Ce-spins coincides with a change in ordered moment and moment fluctuations in the $H-T$ phase space and, hence, suggest that the localized fluctuations can induce a dynamic response to the intrinsic properties of the system.

\begin{figure}
\centering
\includegraphics[width=13.0cm]{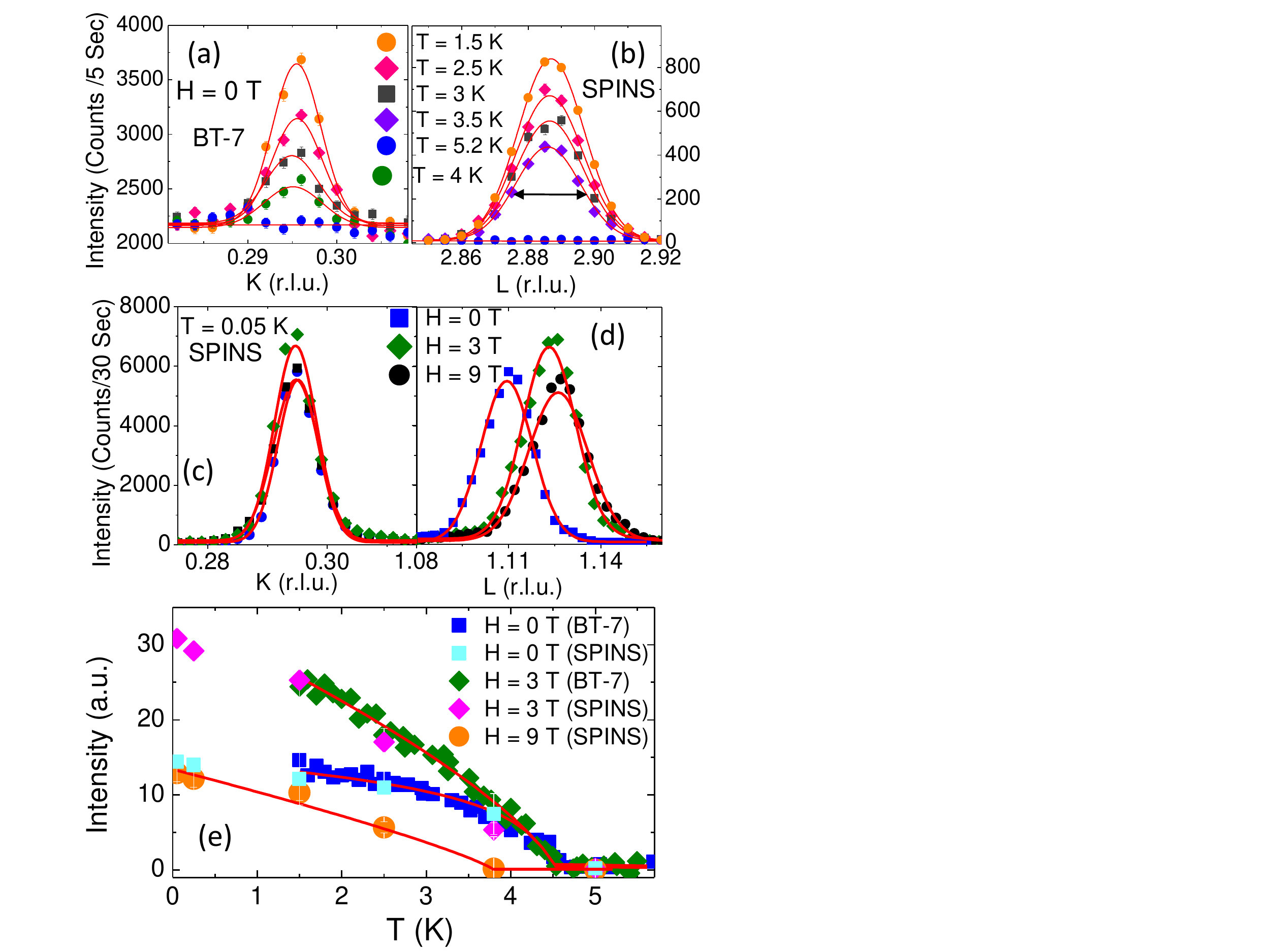} \vspace{-4mm}
\caption{(color online) Elastic scattering scans of the type (a) (0,K,1.11), BT-7 measurements, and
the (b) (0,0.295,L), SPINS measurements, at various temperatures and zero field. (c) and (d) Elastic scans, measured on SPINS, at different fields and $T$= 0.05 K. (e) Intensity at the (0,0.295,1.11) magnetic peak
position at different fields as a function of temperature. It is worth noting that the $l$-value of the magnetic peak position changes with applied field (fig. 1d). The field measurements were performed at the actual magnetic peak position. Error bars represent one standard deviation.
} \vspace{-4mm}
\end{figure}

Neutron scattering measurements were performed on a 0.5 gm single crystal of CeAg$_{2}$Ge$_{2}$ at the spin-polarized triple-axis spectrometer (SPINS) and thermal triple-axis spectrometers (BT7 and BT9) with fixed final neutron energies of 3.7 meV and 14.7 meV, respectively. The high quality of the single crystal was verified by means of XRAY and Laue diffraction pattern.\cite{Thamizhavel} The measurements at SPINS employed a flat pyrolytic graphite (PG) analyzer with collimator sequence Mono-80$^{'}$-sample-BeO-80$^{'}$-flat analyzer-open while the measurements on the thermal spectrometers were performed using a flat PG analyzer with collimations PG filter-Mono-50$^{'}$-sample-PG filter-50$^{'}$-Flat analyzer-120$^{'}$. Measurements were performed with the crystal oriented in the ($0KL$) scattering plane and mounted in a dilution fridge with a base temperature of 50 mK. Here $K$ and $L$ represent reciprocal lattice units of 2${\pi}/b$ and 2${\pi}/c$, respectively. For the magnetic field measurements, the field was applied along the easy [100] direction.

Neutron scattering measurements performed on powder CeAg$_{2}$Ge$_{2}$\cite{Loidl} had identified magnetic reflections with a modulation vector, $\epsilon$, of (0.0,0.285,0.9). Our single-crystal data allow for a much more detailed examination of the intensities and magnetic scattering pattern. In Fig. 1a and 1b, we plot elastic data obtained on BT-7 and SPINS for representative scans at different temperatures along the K- and L- directions. The incommensurate peak positions, defined by $\textbf{q}$ =$\textbf{$\tau$}$$\pm$$\epsilon$ where $\textbf{$\tau$}$ and $\epsilon$ are the nuclear and modulation vectors respectively, are found to be temperature independent at $\textbf{q}$. As the sample is cooled below 4.5 K, resolution limited magnetic peaks (with a lower limit of 144 unit cells) clearly develop indicating the development of long range magnetic order in the system. 

We have performed detailed elastic scattering measurements at $T$ = 1.5 K and $H$ = 0 T in the $(0KL)$-zone spanning a wide region of reciprocal space using 14.7 meV neutrons, as shown in Fig. 2a. All four satellites of the incommensurate magnetic peaks were observed around the nuclear peak positions, while no higher harmonics of incommensurate wave vectors were observed. The best fit of the data is obtained with the Ce-spins lying within the basal plane and forming an amplitude modulated SDW with spins pointing along the $b$-axis. The periodicity of the amplitude modulation is given by the modulation vector $\epsilon$ = (0.0,0.705,0.11) with respect to a simple commensurate structure. The numerically simulated elastic scattering pattern depicted with this spin arrangement is shown in Fig. 2 (b). Since there are four equivalent domains for this spin structure, our calculation averages over all four magnetic domains. We see that the calculated cross-sections are in good agreement with the experimental data. Some of the weaker satellite peaks as found in the magnetic structure calculation were hard to distinguish from the background. Elastic measurements were also performed with the CeAg$_{2}$Ge$_{2}$ crystal rotated in ($HHL$)-zone, but no magnetic Bragg peaks were observed in this plane. We determine that the static moment associated with the long-range incommensurate magnetic peaks at T = 1.5 K is M$_{Ce}$ $\simeq$ 1.23(3) $\mu$$_{B}$, significantly smaller than that expected for a fully degenerate $J$ = 5/2 ground state Ce$^{3+}$ ion value of 2.15 $\mu$$_{B}$. 

\begin{figure}
\centering
\includegraphics[width=13.5cm]{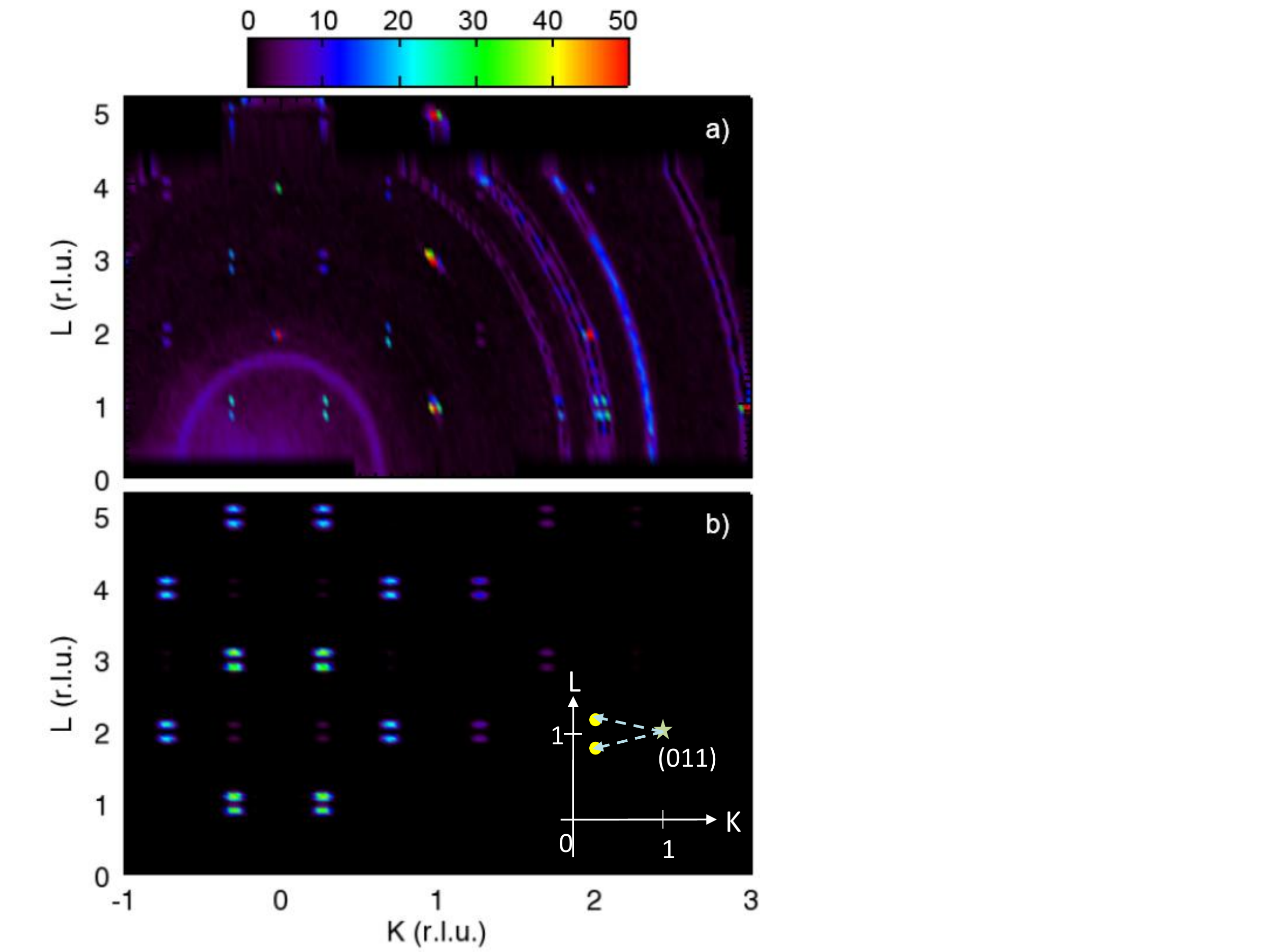} \vspace{-2mm}
\caption{(color online) (a) Pattern of elastic intensity at $T=1.5$~K measured on BT9 in the ($0KL$) zone. The circles at high and low $q$ arise from the Al powder and $^{4}$He, respectively. (b) Calculated
scattering pattern for long-range magnetic order using the
spin density wave magentic configuration, which is domain-averaged as
described in the text. Inset illustrates the modulation wave vector.
} \vspace{-4mm}
\end{figure}

Qualitatively different behavior is observed for the field-dependent scans. Representative scans at $T$ = 0.05 K along K- and L- directions at $\textbf{q}$ = (0.0,0.295,1.11) and at different magnetic fields are plotted in Fig. 1c and 1d. The data clearly reveal the movement of the magnetic peak positions along the [00L] direction. The temperature dependence of the magnetic peak intensity at $\textbf{q}$ = (0.0,0.295,1.11) at different magnetic fields is plotted in Fig. 1e, which reveals that the magnetic intensity at 3 T is higher compared to 0 T or 9 T data. This suggests a fluctuation in the ordering moment as a function of field. Corrections were made in the $l$-value for the field measurements such that $\textbf{q}$ reperesents the actual magnetic peak position. Fitting the data to a power-law, $I$ $\propto$ (1-$T$/$T$$_{N}$)$^{-2\beta}$, yields the AFM transition temperature $T_N \simeq 4.5$~ K at $H$ = 0 and 3 T. At $H$= 9 T, $T_N$ shifts to lower temperature ($\simeq$~3.8 K). The exponent, $\beta$, at 0 and 3 T is 0.28 (2) and 0.34 (2) respectively, suggesting 3-D interactions.\cite{Andrea}

\begin{figure}
\centering
\includegraphics[width=10.8cm]{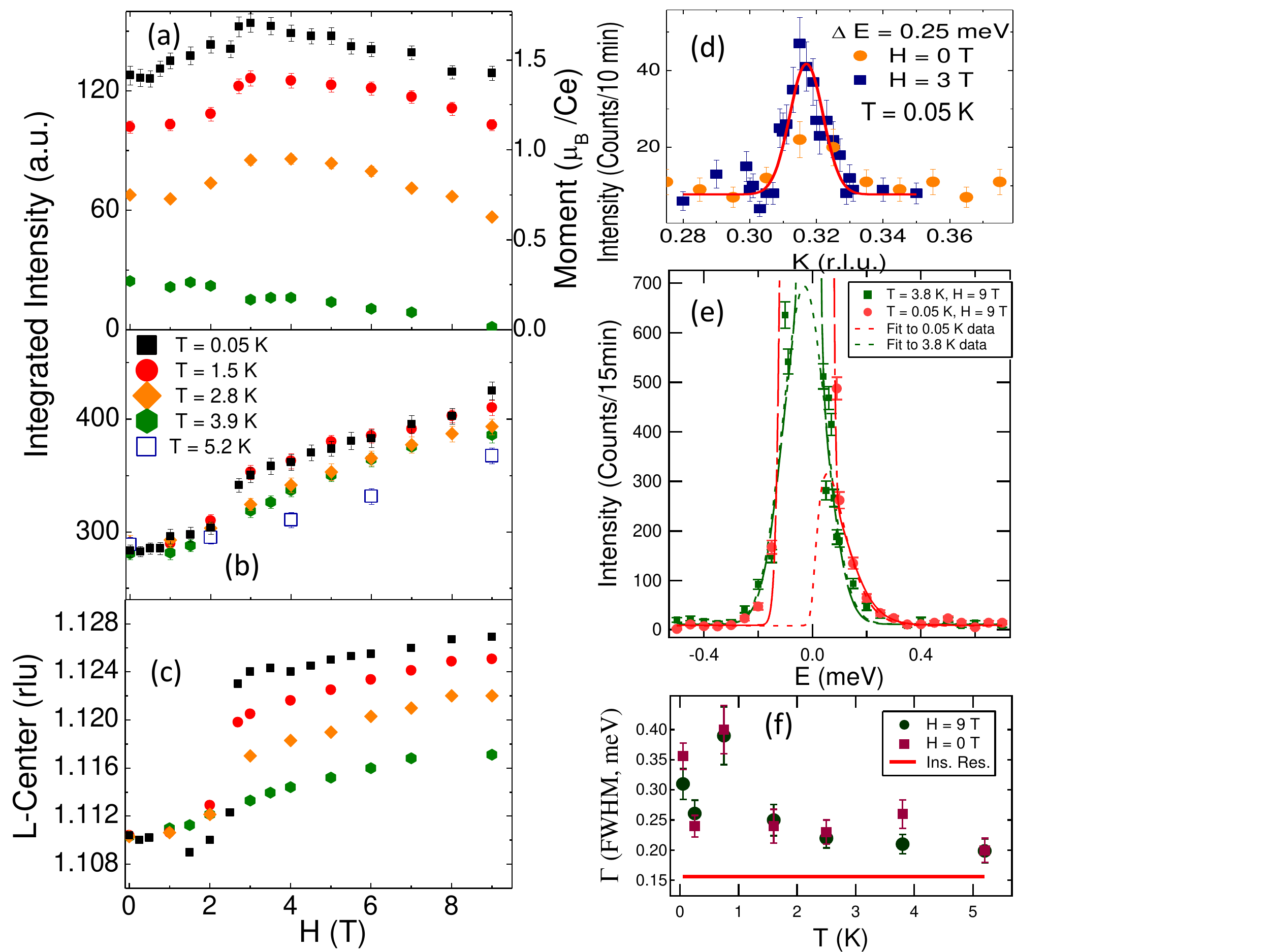} \vspace{-4mm}
\caption{(color online) (a) and (b) Variation of integrated magnetic intensity at (0,0.295,1.11) and nuclear intensity at (0,0,2) as a function of field at different temperatures. (c) Change in L-center position of magnetic Bragg peak at (0,0.295,L) as a function of field. An abrupt change is clearly observed at $H$ = 2.7 T. (d) Constant energy scan at $E$ = 0.25 meV along $K$-axis at (0,K,1.115) clearly shows enhanced fluctuation at $H$ = 3 T field, which is when there is an abrupt change in (a) and (c) as well. (e) Representative inelastic scans at two temperatures and 9 T. (f) Temperature dependence of the observed linewidths (FWHM) at $H$ = 0 and 9 T. The incoherent scattering from vanadium was used to determine the energy resolution.
} \vspace{-4mm}
\end{figure}

In Fig. 3a and 3b, we have plotted the field-dependent scans of integrated intensities as a function of temperature at magnetic peak position $\textbf{q}$ = (0.0,0.295,1.11) and the nuclear peak position $\textbf{$\tau$}$ = (0,0,2), respectively. We see that the integrated intensity of the magnetic Bragg peak starts increasing as the applied field is increased to 2.7 T, develops a maxima around 3 T field followed by a gradual decrease as the applied field value is increased to 9 T. In terms of the ordered moment, it changes from 1.43 $\mu_B$ at $H$ = 0 T-to-1.61 $\mu_B$ at $H$ = 2.7 T-to-1.4 $\mu_B$ at $H$ = 9 T at $T$=0.05 K. This behavior in field continues until $T$$\simeq$3.8 K, while for $T$ $\geq$3.8 K the integrated magnetic intensity continuously decreases as the field is increased. Similar patterns in the change of magnetic intensities were observed at other magnetic peak positions. Interestingly, there is a substantial induced magnetic moment (0.23 $\mu_B$ per Ce-ion at $H$ = 3 T at $T$=0.05 K) observed as additional scattering on the (002) nuclear peak which monotonically increases with fields. The intensity at the (002) position also increases above $T_N$$\simeq$4.5 K as a function of field, therefore suggesting that the observed behaviors at magnetic Bragg peaks and (00L) magneto-nuclear peaks are decoupled. Measurements were also performed for reverse field sweeps and no hysteresis was observed in magnetic or nuclear case. The increase in the (002) peak intensity in field is either due to the field-induced alignment of free Ce-moments parallel to the field or, the induced moment from the intermixing of lower level crystal field excitations in applied filed or, a combination of both.\cite{Lussier} More research works are needed to understand that.

The ordered moment of correlated Ce-ions, forming the SDW configuration, exhibits subtle changes as a function of field below $T$$\leq$3.8K. In a notable observation, as shown in Fig. 3c, we have found that the onset of moment variation at $H$=2.7 T is accompanied by a discontinuous movement of the SDW propagation vector along the $\textbf{L}$ direction. In reciprocal space, the $l$ value of the magnetic vector changes from $l$=1.11 at $H$=0 T to $l$~=1.123 at $H$=2.7 T and $T$=0.05 K. As the applied field is increased further ($H$$\geq$~5 T), the movement of the SDW moderates and tends to settle at higher field values. This coincidence of the change in ordered moment of Ce-ions and the discontinuous movement of the underlying SDW in $H-T$ space clearly suggests a coupling between them. At higher fields, the magnetic intensity decreases at all temperatures, which is consistent with the magnetic behavior of AFM materials. No such behavior was observed at the magneto-nuclear peak position (002). 

Several possibilities were explored to explain this unusual coupling, including the phenomena of field-induced intermixing of crystal field levels. Previous measurements\cite{Loidl,Thamizhavel} of CeAg$_{2}$Ge$_{2}$ suggested that the crystal field levels of Ce-ions consists of a doublet ground state, either degenerate with a doublet first excited state or separated by $\simeq$ 5 K. Therefore the effect of magnetic field perturbation resulting into the intermixing of crystal field levels cannot be ruled out. However, the robust nature of SDW magnetic structure of Ce ions and the absence of higher harmonics or, lock-in phase at low temperature do not suggest that. Field-induced anti-ferroquadrupolar (AFQ)-type magnetic order in doublet ground state systems has been observed before but they are usually accompanied with higher order harmonics at lower temperature. Also, such tendency is clearly observed in heat capacity measurements.\cite{Onimaru} As shown below in Fig.4, no such anomaly is observed in the heat capacity measurements of CeAg$_{2}$Ge$_{2}$ at $H$$\simeq$3 T.

We also performed inelastic measurements. In Fig. 3d, we have plotted constant energy scans along (0,$K$,1.115) direction at $E$=0.25 meV at $H$=0 and 3 T. Significantly higher intensity is observed at 3 T compared to 0 T data. Fluctuations can lead to the increase or decrease in the magnetic intensity. Resolution limited Gaussian fit of the 3 T data indicate the dynamic nature of Ce moments with long range spatial correlation. Representative scans of background corrected inelastic measurements between $T$=0.05 K and 5.5 K and $H$=0 and 9 T at the AFM wave vector, $\textbf{q}$ = (0,0.295,1.11) are shown in Fig. 3e. A combination of two Gaussian profiles, representing elastic and quasi-elastic lines respectively, multiplied by the detailed balance factor and convoluted with the instrument resolution gives a better fit of the experimental data. At T=0.05 K, the quasi-elastic line shifts to slightly inelastic position at $E$=0.1 meV. In Fig. 3f, we have plotted the estimated linewidths (FWHM) at different temperatures along with the instrument resolution at $H$ = 0 and 9 T. While the estimated linewidths are found to be reasonably broader, $\simeq$ 0.25 meV, than the instrument resolution (0.158 meV), no significant temperature dependence is observed below $T$ $\leq$3.8 K. The Gaussian approximation of quasi-elastic data suggests correlated AFM fluctuation of Ce-ions, as reported previously in the case of the quantum critical AFM compound U$_{2}$Zn$_{17}$.\cite{Walter} The very weak temperature dependence of the linewidths possibly indicates the dominant effect of crystal-field coupled intersite exchagne coupling between Ce-ions.\cite{Broholm} More work is needed to fully understand this behavior. 

\begin{figure}
\centering
\includegraphics[width=8.0cm]{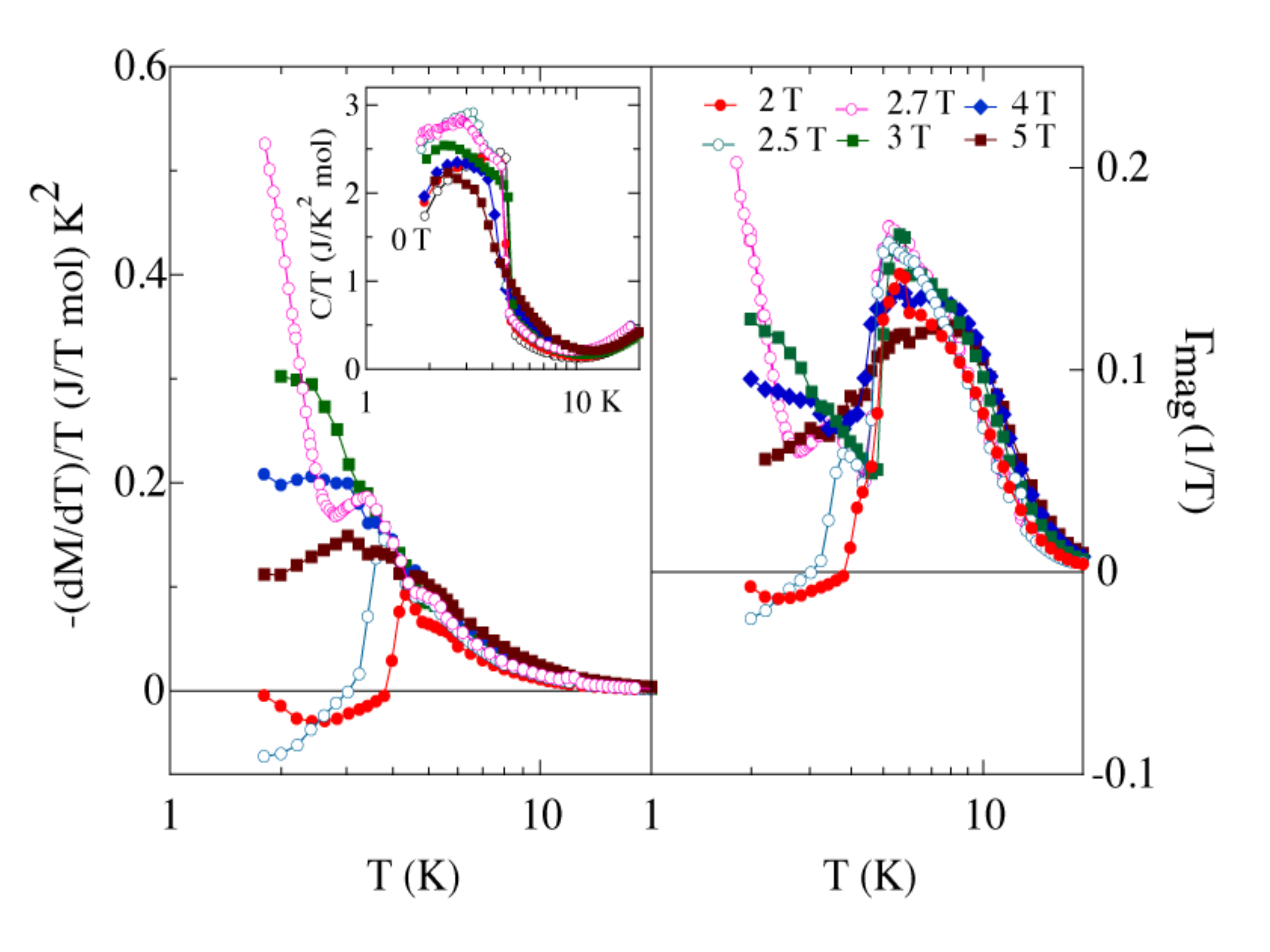} \vspace{-2mm}
\caption{(color online) (a) Differential magnetic susceptibility at different fields as a function of temperature. C(T)/$T$ at different fields as a function of temperature are shown in the inset. (b) Variation of Gruneisen ratio parameter at different fields. 
} \vspace{-4mm}
\end{figure}

Finally, we have performed magnetic Gruneisen ratio measurements to gain more insight into the nature of Ce moment fluctuations in the $H-T$ phase space. The divergence of the Gruneisen ratio parameter defined by the ratio of -($d$M/$d$T)/$T$ and C(T)/$T$, such that $\Gamma$$_{mag}$= -($d$M/$d$T)/$C$, is described as an apt way to check the spin fluctuations.\cite{Steglich,Zhu} In Fig. 4, we plot the temperature dependence of -($d$M/$d$T)/$T$ (Fig. 4a), C(T)/$T$ (inset, Fig. 4a) and $\Gamma$$_{mag}$ in Fig. 4b. While the magnetization was measured utilizing a high resolution commercial magnetometer, heat capacity measurements were performed using the relaxation method in a commercial physical properties measurement system with magnetic field applied along [100] direction. Two observations are clearly noticeable in this figure: (a) a clear divergence in the ratio $\Gamma$$_{mag}$ at $H$=2.7 T, dominated by the divergence in differential susceptibility, and (b) changes in the slope of $\Gamma$$_{mag}$ curves across $H$=2.7 T and $T$$\leq$3.8 K. Although, the heat capacity ratio increases sharply at $T$$\sim$~4.5 K, confirming the magnetic phase transition at $T$$_{N}$ = 4.5 K, the lack of strong field dependence of C(T)/$T$ in the explored field regime suggests that the electronic density is weakly affected by applied magnetic field in CeAg$_{2}$Ge$_{2}$. Therefore, the observed linear divergence of $\Gamma$$_{mag}$ at $H$ = 2.7 T and the changes in the slope of the curves across this field value are magnetic in origin, which are in conjunction with the field-dependent neutron scattering measurements (Fig. 3).

In summary, we have observed a significant change in ordered Ce moments as a function of field. The onset of this behavior coincides with the discontinuous movement of intrinsic SDW along $L$ and significant fluctuations along the $K$-direction. It is not yet clear if the observed fluctuation of Ce-moments will lead to a quantum phase transition at $T$= 0 K. We point out that the sliding of the SDW is in essence an intrinsic elastic phenomena.\cite{Eric} However, the unusual discontinuous movement of the SDW in CeAg$_{2}$Ge$_{2}$ suggests a dynamic behavior, accompanied by a finite change in total magnetic energy, which tends to compensate the Ce fluctuations by spatially changing the propagation vector. Previously, Landau had described that a discontinous change in the magnetic vector involves a change in total magnetic energy.\cite{Landau} Further understanding of this interesting behavior will require detailed theoretical calculations based on the total energy across the field boundary. 

This work used facilities supported in part by the NSF under Agreement No. DMR-0944772.

\bibliography{CAG}

\begin{thebibliography}{22}
\expandafter\ifx\csname natexlab\endcsname\relax\def\natexlab#1{#1}\fi
\expandafter\ifx\csname bibnamefont\endcsname\relax
  \def\bibnamefont#1{#1}\fi
\expandafter\ifx\csname bibfnamefont\endcsname\relax
  \def\bibfnamefont#1{#1}\fi
\expandafter\ifx\csname citenamefont\endcsname\relax
  \def\citenamefont#1{#1}\fi
\expandafter\ifx\csname url\endcsname\relax
  \def\url#1{\texttt{#1}}\fi
\expandafter\ifx\csname urlprefix\endcsname\relax\def\urlprefix{URL }\fi
\providecommand{\bibinfo}[2]{#2}
\providecommand{\eprint}[2][]{\url{#2}}

\bibitem[{\citenamefont{Gegenwart\emph{,et al.}}(2008)}]{Steglich}
\bibinfo{author}{\bibfnamefont{P.}~\bibnamefont{Gegenwart\emph{,et al.}}},
  \bibinfo{journal}{Nature Physics} \textbf{\bibinfo{volume}{4}},
  \bibinfo{pages}{186} (\bibinfo{year}{2008}).

\bibitem[{\citenamefont{Si et~al.}(2001)\citenamefont{Si, Rabello, Ingersent,
  and Smith}}]{Si}
\bibinfo{author}{\bibfnamefont{Q.}~\bibnamefont{Si}},
  \bibinfo{author}{\bibfnamefont{S.}~\bibnamefont{Rabello}},
  \bibinfo{author}{\bibfnamefont{K.}~\bibnamefont{Ingersent}},
  \bibnamefont{and} \bibinfo{author}{\bibfnamefont{J.~L.} \bibnamefont{Smith}},
  \bibinfo{journal}{Nature} \textbf{\bibinfo{volume}{413}},
  \bibinfo{pages}{804} (\bibinfo{year}{2001}).

\bibitem[{\citenamefont{Lohneysen\emph{,et al.}}(2007)}]{Lohneysen}
\bibinfo{author}{\bibfnamefont{H.}~\bibnamefont{Lohneysen\emph{,et al.}}},
  \bibinfo{journal}{Rev. Mod. Phys.} \textbf{\bibinfo{volume}{79}},
  \bibinfo{pages}{1015} (\bibinfo{year}{2007}).

\bibitem[{\citenamefont{Coleman\emph{,et al.}}(2001)}]{Coleman}
\bibinfo{author}{\bibfnamefont{P.}~\bibnamefont{Coleman\emph{,et al.}}},
  \bibinfo{journal}{J. Phys.: Cond. Matt.} \textbf{\bibinfo{volume}{13}},
  \bibinfo{pages}{R723} (\bibinfo{year}{2001}).

\bibitem[{\citenamefont{Hertz}(1976)}]{Hertz}
\bibinfo{author}{\bibfnamefont{J.~A.} \bibnamefont{Hertz}},
  \bibinfo{journal}{Phys. Rev. B} \textbf{\bibinfo{volume}{14}},
  \bibinfo{pages}{1165} (\bibinfo{year}{1976}).

\bibitem[{\citenamefont{Mills}(1993)}]{Millis}
\bibinfo{author}{\bibfnamefont{A.~J.} \bibnamefont{Mills}},
  \bibinfo{journal}{Phys. Rev. B} \textbf{\bibinfo{volume}{48}},
  \bibinfo{pages}{7183} (\bibinfo{year}{1993}).

\bibitem[{\citenamefont{Moriya and Takimoto}(1995)}]{Moriya}
\bibinfo{author}{\bibfnamefont{T.}~\bibnamefont{Moriya}} \bibnamefont{and}
  \bibinfo{author}{\bibfnamefont{T.}~\bibnamefont{Takimoto}},
  \bibinfo{journal}{J. Phys. Sco. Jpn.} \textbf{\bibinfo{volume}{64}},
  \bibinfo{pages}{960} (\bibinfo{year}{1995}).

\bibitem[{\citenamefont{Knafo\emph{, et al.}}(2009)}]{Knafo}
\bibinfo{author}{\bibfnamefont{W.}~\bibnamefont{Knafo\emph{, et al.}}},
  \bibinfo{journal}{Nature Physics} \textbf{\bibinfo{volume}{5}},
  \bibinfo{pages}{753} (\bibinfo{year}{2009}).

\bibitem[{\citenamefont{Kadowaki\emph{, et al.}}(2006)}]{Kadowaki}
\bibinfo{author}{\bibfnamefont{H.}~\bibnamefont{Kadowaki\emph{, et al.}}},
  \bibinfo{journal}{Phys. Rev. Lett.} \textbf{\bibinfo{volume}{96}},
  \bibinfo{pages}{016401} (\bibinfo{year}{2006}).

\bibitem[{\citenamefont{Aronson\emph{, et al.}}(1995)}]{Aronson}
\bibinfo{author}{\bibfnamefont{M.~C.} \bibnamefont{Aronson\emph{, et al.}}},
  \bibinfo{journal}{Phys. Rev. Lett.} \textbf{\bibinfo{volume}{75}},
  \bibinfo{pages}{725} (\bibinfo{year}{1995}).

\bibitem[{\citenamefont{Stockert\emph{, et al.}}(2007)}]{Stockert}
\bibinfo{author}{\bibfnamefont{O.}~\bibnamefont{Stockert\emph{, et al.}}},
  \bibinfo{journal}{Phys. Rev. Lett.} \textbf{\bibinfo{volume}{99}},
  \bibinfo{pages}{237203} (\bibinfo{year}{2007}).

\bibitem[{\citenamefont{Wilson\emph{, et al.}}(2005)}]{Wilson}
\bibinfo{author}{\bibfnamefont{S.~D.} \bibnamefont{Wilson\emph{, et al.}}},
  \bibinfo{journal}{Phys. Rev. Lett.} \textbf{\bibinfo{volume}{94}},
  \bibinfo{pages}{056402} (\bibinfo{year}{2005}).

\bibitem[{\citenamefont{Loidl\emph{, et al.}}(1992)}]{Loidl}
\bibinfo{author}{\bibfnamefont{A.}~\bibnamefont{Loidl\emph{, et al.}}},
  \bibinfo{journal}{Phys. Rev. B} \textbf{\bibinfo{volume}{46}},
  \bibinfo{pages}{9341} (\bibinfo{year}{1992}).

\bibitem[{\citenamefont{Thamizhavel\emph{, et al.}}(2007)}]{Thamizhavel}
\bibinfo{author}{\bibfnamefont{A.}~\bibnamefont{Thamizhavel\emph{, et al.}}},
  \bibinfo{journal}{Phys. Rev. B} \textbf{\bibinfo{volume}{75}},
  \bibinfo{pages}{144426} (\bibinfo{year}{2007}).

\bibitem[{\citenamefont{Pelissetto and Vicari}(2002)}]{Andrea}
\bibinfo{author}{\bibfnamefont{A.}~\bibnamefont{Pelissetto}} \bibnamefont{and}
  \bibinfo{author}{\bibfnamefont{E.}~\bibnamefont{Vicari}},
  \bibinfo{journal}{Physics Reports} \textbf{\bibinfo{volume}{368}},
  \bibinfo{pages}{549} (\bibinfo{year}{2002}).

\bibitem[{\citenamefont{Lussier\emph{, et al.}}(1997)}]{Lussier}
\bibinfo{author}{\bibfnamefont{J.~G.} \bibnamefont{Lussier\emph{, et al.}}},
  \bibinfo{journal}{Phys. Rev. B} \textbf{\bibinfo{volume}{56}},
  \bibinfo{pages}{11749} (\bibinfo{year}{1997}).

\bibitem[{\citenamefont{Onimaru\emph{, et al.}}(2005)}]{Onimaru}
\bibinfo{author}{\bibfnamefont{T.}~\bibnamefont{Onimaru\emph{, et al.}}},
  \bibinfo{journal}{Phys. Rev. Lett.} \textbf{\bibinfo{volume}{94}},
  \bibinfo{pages}{197201} (\bibinfo{year}{2005}).

\bibitem[{\citenamefont{Walter\emph{, et al.}}(1987)}]{Walter}
\bibinfo{author}{\bibfnamefont{U.}~\bibnamefont{Walter\emph{, et al.}}},
  \bibinfo{journal}{Phys. Rev. B} \textbf{\bibinfo{volume}{36}},
  \bibinfo{pages}{1981} (\bibinfo{year}{1987}).

\bibitem[{\citenamefont{Broholm\emph{,et al.}}(1987)}]{Broholm}
\bibinfo{author}{\bibfnamefont{C.}~\bibnamefont{Broholm\emph{,et al.}}},
  \bibinfo{journal}{Phys. Rev. Lett.} \textbf{\bibinfo{volume}{58}},
  \bibinfo{pages}{917} (\bibinfo{year}{1987}).

\bibitem[{\citenamefont{Zhu\emph{, et al.}}(2003)}]{Zhu}
\bibinfo{author}{\bibfnamefont{L.}~\bibnamefont{Zhu\emph{, et al.}}},
  \bibinfo{journal}{Phys. Rev. Lett.} \textbf{\bibinfo{volume}{91}},
  \bibinfo{pages}{066404} (\bibinfo{year}{2003}).

\bibitem[{\citenamefont{Fawcett}(1988)}]{Eric}
\bibinfo{author}{\bibfnamefont{E.}~\bibnamefont{Fawcett}},
  \bibinfo{journal}{Rev. Mod. Phys.} \textbf{\bibinfo{volume}{60}},
  \bibinfo{pages}{209} (\bibinfo{year}{1988}).

\bibitem[{\citenamefont{Landau}(1937)}]{Landau}
\bibinfo{author}{\bibfnamefont{L.~D.} \bibnamefont{Landau}},
  \bibinfo{journal}{Phys. Z. Sowjetunim} \textbf{\bibinfo{volume}{11}},
  \bibinfo{pages}{26} (\bibinfo{year}{1937}).

\end{thebibliography}
\end{document}